\documentclass[lettersize,journal]{IEEEtran}
\usepackage{amsmath,amsfonts}
\usepackage{algorithmic}
\usepackage{algorithm}
\usepackage{array}
\usepackage{textcomp}
\usepackage{stfloats}
\usepackage{url}
\usepackage{verbatim}
\usepackage{graphicx}
\usepackage{cite}

\usepackage{hyperref,lineno,microtype,subcaption}
\usepackage{soul,xcolor}
\usepackage{adjustbox}
\begin{document}

\title{Predicting Side Effect of Drug Molecules \\ using Recurrent Neural Networks \vspace{-4mm}}
\author{Collin Beaudoin$^{1}$, Koustubh Phalak$^{1}$, Swaroop Ghosh$^{1}$

$^{1}$Penn State University, Department of Computer Science and Engineering, University Park, Pennsylvania, USA }



\maketitle
\vspace{-16mm}
\begin{abstract}
Identification of molecular properties, like side effects, is one of the most important and time-consuming steps in the process of molecule synthesis. Failure to identify side effects before submission to regulatory groups can cost millions of dollars and months of additional research to the companies. Failure to identify side effects during the regulatory review can also cost lives. The complexity and expense of this task have made it a candidate for a machine learning-based solution. Prior approaches rely on complex model designs and excessive parameter counts for side effect predictions. Reliance on complex models only shifts the difficulty away from chemists rather than alleviating the issue. Implementing large models is also expensive without prior access to high-performance computers. We propose a heuristic approach that allows for the utilization of simple neural networks, specifically the GRU recurrent neural network, with a 98+\% reduction of required parameters compared to available large language models while obtaining near identical results as top-performing models.
\end{abstract}
\vspace{-2mm}

\footnote{This work has been submitted to the IEEE for possible publication. Copyright may be transferred without notice, after which this version may no longer be accessible.}

\vspace{-4mm}
\section{Introduction}
\vspace{-2mm}
Molecular property prediction is one of the most fundamental tasks within the field of drug discovery \cite{yang2019analyzing, wieder2020compact}. Applying in silico methods to molecular property prediction offers the potential of releasing safer drugs to the market while reducing test time and cost. Detecting molecular properties before development enables researchers to develop more effective new materials faster and with higher certainty. Detecting known causes of side effects in drugs before release can prevent unnecessary injury and save thousands of lives. \color{black} Historically, these in silico approaches relied on complex feature engineering methods to generate their molecule representations for processing \cite{lengauer2004novel, merkwirth2005automatic}. The bias of the descriptors limits these approaches, which means the generated features may not be reusable for different tasks as some valuable identifiers may not be present. The feature vectors also depend on current molecular comprehension; upon discovery, the feature vectors could become redundant. Graph Neural Networks (GNN) remove the dependence on complex and temporal descriptors. GNNs became favorable due to the common practice of drawing molecules using graph representations, which offer a generic form for the input. The generic input format allows machine learning models to build their interpretation of information rather than rely on human capabilities. Through these advances GNNs perform well on multiple chem-informatic tasks, especially molecular property prediction \cite{hu2019strategies, wu2020comprehensive}. Despite these improvements GNNs still have limitations. Specifically, GNNs have difficulty understanding shared dependence and have scalability issues. The size of the graphical input increases exponentially with each additional molecule that is represented. With this growth, the cost of communication between graphical nodes also exponentially increases. Compared to other neural network types, GNNs can perform worse at molecular property prediction, despite their built-in generic representation \cite{mayr2018large}. With the recent success of large language models, newer attempts aim to build transformer-based approaches with promising signs of success \cite{taylor2022galactica}. While new large language models offer comparable performance to GNNs, they require up to 120 billion parameters. 

Due to the rapid explosion of parameters caused by GNNs, feed-forward neural networks, and transformers, we propose a heuristic approach using a recurrent neural network, specifically the gated recurrent unit (GRU)\color{black}. Our approach can obtain close to state-of-the-art results with 99+\% fewer parameters than large graph-based models or large language-based models, such as \color{black} Galactica \cite{taylor2022galactica}. In the following sections, we review the MoleculeNet \color{black} benchmark \cite{wu2018moleculenet} and compare the SMILES and SELFIES formats and the basic concepts of a recurrent neural network, and also discuss a few of the related works that are evaluated using the MoleculeNet benchmark \color{black} (Section II). We then discuss the data pre-processing and model implementation details (Section III), followed by model performances and a comparison to other state-of-the-art options (Section IV). Finally, we conclude the paper by giving a summary (Section V). 

\vspace{-6mm}
\section{Background \& Related Works}

\subsection{MoleculeNet Benchmark} \label{molnet_background}
\vspace{-2mm}

MoleculeNet is a benchmark set used to evaluate machine learning techniques \cite{wu2018moleculenet}. It curate's quantum mechanics, physical chemistry, biophysics, and physiology datasets. For each dataset, it establishes the preferred metric for evaluation to enable consistent comparison across models. We describe each dataset selected to evaluate our model.

\subsubsection{Side Effect Resource (SIDER)}
The principal molecular property of human consumption is the side effects associated with the molecule. The Side Effect Resource (SIDER) dataset attempts to create a single source of combined public records for known side effects \cite{kuhn2016sider}. The dataset consists of 28 columns; the first column is the SMILES representation of a given molecule, and the 27 subsequent columns are affected system organ classes where side effects are classified by MedDRA \footnote{https://www.meddra.org/}. The side effects of each molecule are marked with a one if it is known to have a side effect or a zero otherwise. 

\subsubsection{BACE} 
BACE is a collection of experimentally reported values from various journals for the binding results for inhibitors of human $\beta$-secretase 1 \cite{subramanian2016computational}.

\subsubsection{Blood-brain barrier penetration (BBBP)}
Here molecules are classified by their ability to permeate through the blood-brain barrier. A drug’s ability to permeate through the blood barrier is an important feature for drugs specifically targeting the central nervous system \cite{martins2012bayesian}.

\subsubsection{ClinTox}
MoleculeNet introduces ClinTox to evaluate drugs previously approved by the FDA and drugs that have failed clinical trials due to toxicity.  \cite{wu2018moleculenet}

\subsubsection{HIV}
The HIV dataset is originally from the Drug Therapeutics Program (DTP) \footnote{https://wiki.nci.nih.gov/display/NCIDTPdata/AIDS+Antiviral+Screen+Data} consisting of molecules tested to inhibit HIV replication. There are roughly 40k samples within the dataset, where MoleculeNet uses two labels, confirmed inactive and confirmed active. 

\subsubsection{MUV}
The Maximum Unbiased Validation (MUV) dataset contains 17 labeling tasks and 90k molecules. The dataset originates from PubChem \cite{rohrer2009maximum}.

\vspace{-6mm}
\subsection{ROC-AUC}
\vspace{-2mm}
The receiver operating characteristic curve (ROC curve) measures the true positive rate against the false positive rate at multiple threshold settings for a binary classifier. This measures the ability of a model to distinguish correctly between two classes. ROC-AUC is commonly preferred when evaluating models trained on imbalanced datasets, making it an ideal statistic to evaluate the MoleculeNet datasets.
\color{black}

\vspace{-6mm}
\subsection{Simplified Molecular-Input Line Entry System (SMILES)}
Simplified molecular-input line-entry system (SMILES) uses characters to build a molecular representation \cite{weininger1988smiles}. Letters represent various elements within a molecule, where the first letter of an element can be uppercase, denoting that the element is non-aromatic, or lowercase, denoting that the element is aromatic. Assuming an element requires a second letter, it will be lowercase. Another possible representation of aromaticity is the colon, which is the aromatic bond symbol. Other potential bond symbols are a period (.), a hyphen (-), a forward slash (/), a backslash (\textbackslash), an equal sign (=), an octothorpe (\#), and a dollar sign (\$). Periods represent a no bond, hyphens represent a single bond, and the forward slash and backslash represent single bonds adjacent to a double bond. However, the forward slash and backslash are only necessary when rendering stereochemical molecules. The equal sign represents a double bond, the octothorpe represents the triple bond, and the dollar sign represents a quadruple bond. In cases where stereochemical molecules are used, the asperand (@) can be used in a double instance to represent clockwise or in a single occurrence to represent counterclockwise. Numbers are used within a molecule to characterize the opening and closing of a ring structure, or if an element is within brackets, the number can represent the number of atoms associated with an element. Numbers appearing within brackets before an element represent an isotope. A parenthesis (()) denotes branches from the base chain.

\vspace{-4mm}
\subsection{Self-Referencing Embedded Strings (SELFIES)}

Self-referencing embedded Strings (SELFIES) improve the initial idea of SMILES for usage in machine learning processes by creating a robust molecular string representation \cite{krenn2020self}. SMILES offered a simple and interpretable characterization of molecules that was able to encode the elements of molecules and their spatial features. The spatial features rely on an overly complex grammar where rings and branches are not locally represented features. This complexity causes issues, especially in generative models, where machines frequently produce syntactically invalid or physically invalid strings. To remove this non-locality, SELFIES uses a single ring or branch symbol, and the length of this spatial feature is directly supplied; ensuring that any SELFIES string has a valid physical representation.

\vspace{-4mm}
\subsection{Recurrent Neural Networks (RNN)}
\begin{figure}[t]
\vspace{-4mm}
  \centering
  \includegraphics[width=0.7\linewidth]{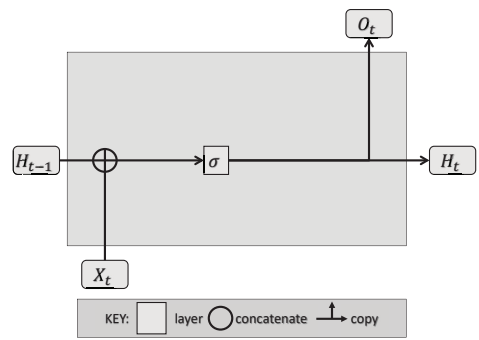}
  \caption{Vanilla RNN architecture used for training; ($H_{t-1}, H_t$) represent the hidden state, ($O_{t}$) represents the output state, and ($X_t$) represents the input information. The $\sigma$ represents the activation function that operates on the combined input and hidden state. }   \label{fig:RNN}
  \vspace{-8mm}
\end{figure}

\vspace{-2mm}
Elman networks, more commonly known as vanilla recurrent neural networks (RNN), attempt to introduce the concept of a time-dependent dynamic memory \cite{elman1990finding}. The idea is to make predictions about inputs based on contextual information. Context-based predictions can be done for four input-output schemes: one-to-one, one-to-many, many-to-one, and many-to-many. One-to-one models are a variation of a classic neural network, one-to-many models are best for image caption generation, many-to-one models are best for sentiment analysis, and many-to-many models are best for translation or video frame captioning. Fig. ~\ref{fig:RNN} is an example of the basic structure of a vanilla RNN.

In Fig. ~\ref{fig:RNN}, the $X_t$ represents some input, $H_{t-1}, H_t$ represents some hidden state (which is representative of memory), $O_t$ represents some output, and $\sigma$ represents some activation function. The current input information combines with the previous hidden state, and the resulting combined state is then fed to an activation function to insert some non-linearity. This non-linearity produces the next hidden state, which can be manipulated to create a desired output. The fundamental element is the hidden state. The hidden state theoretically allows for consideration of any historical input and its effects on the current input. For a mathematical description of an RNN, we refer to Equation ~\ref{eq:H_t} and Equation ~\ref{eq:O_t}.

\vspace{-4mm}
\begin{equation}
H_{t} = \sigma(W_{HH} H_{t-1} + W_{XH} X_{t}) \label{eq:H_t}
\end{equation}
\vspace{-4mm}
\begin{equation}
O_{t} = W_{HO}H_{t} \label{eq:O_t}
\end{equation}

\begin{figure}[t]
\vspace{-2mm}
  \centering
  \includegraphics[width=0.7\linewidth]{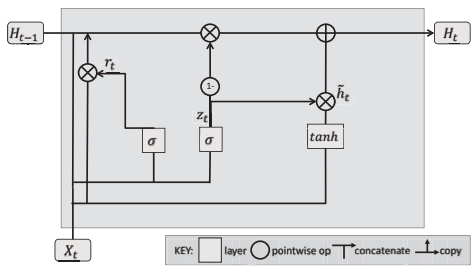}
  \caption{GRU architecture used for training; ($h_{t-1}, h_t$) represent the hidden state, ($\Tilde{h}_{t}$) represents the candidate hidden state state, and ($r_t$) and ($Z_t$) represents the parameters to tune the importance of the previous hidden state versus the updated information. The $\sigma$ represents the activation function that operates on the combined input and hidden state. }   \label{fig:GRU}
  \vspace{-6mm}
\end{figure}

Unfortunately, Vanilla RNNs suffer from memory saturation issues, so they are not always reliable. There have been many methods proposed to overcome this issue, but one of the most popular is the Gated Recurrent Unit (GRU)\cite{cho2014learning}. The basic structure of a GRU is in Fig. ~\ref{fig:GRU}. We can mathematically describe each of the components using Equation ~\ref{eq:Z_t}, Equation ~\ref{eq:r_t}, Equation ~\ref{eq:gru_hold_t}, and Equation ~\ref{eq:gru_h_t}. Equation ~\ref{eq:gru_hold_t} represents the candidate hidden state function, representing the potential updated state. Equation ~\ref{eq:gru_h_t} performs the actual update to the hidden state based on the previous hidden state and the candidate hidden state. Both Equation ~\ref{eq:Z_t} and Equation ~\ref{eq:r_t} allow the network to tune the importance of the contribution of the previous hidden state to the new hidden state. Because of the $r_t$ and $z_t$ parameters, the GRU can better control its memory state offering a practical improved performance over RNNs.

\vspace{-2mm}
\begin{equation}
z_{t} = \sigma(W_{z} \cdot [h_{t-1}, x_{t}] ) \label{eq:Z_t}
\end{equation}
\vspace{-4mm}
\begin{equation}
r_{t} = \sigma(W_{r} \cdot [h_{t-1}, x_{t}]) \label{eq:r_t}
\end{equation}
\vspace{-4mm}
\begin{equation}
\Tilde{h}_{t} = tanh(W \cdot [r_{t}*h_{t-1}, x_{t}]) \label{eq:gru_hold_t}
\end{equation}
\vspace{-4mm}
\begin{equation}
h_{t} = (1-z_{t}) * h_{t-1} + z_{t} * \Tilde{h}_{t} \label{eq:gru_h_t}
\end{equation}
\vspace{-6mm}
\color{black}

\vspace{-4mm}
\subsection{Related Works}

\paragraph{GROVER}
The graph representation from the self-supervised message passing transformer (GROVER) model takes two forms, GROVER$_{base}$ and GROVER$_{large}$ \cite{rong2020self}. We only consider GROVER$_{large}$ as it achieves the highest performance of the two. GROVER bases its design on popular large language models such as, BERT and GPT, where a large corpus of information pre-trains a model and fine-tuning is applied for the completion of downstream tasks \cite{devlin2018bert, radford2018improving}. However, they stray from prior works that attempt training using the SMILES string format \cite{wang2019smiles} and instead use graphs, which they state are more expressive. Previous graph pre-training approaches use the available supervised labels to train their model \cite{hu2019strategies}, but GROVER prefers a self-supervised approach to achieve higher performance, so they suggest using contextual property prediction and graph-level motif prediction. Contextual property prediction takes a given element (node) within a molecular graph and predicts the connected elements and the type of bond used for the connection. Graph-level motif prediction takes a given molecule and attempts to predict the recurrent sub-graphs, known as motifs, that may appear within the molecule. To build the model, they designed a new architecture known as the GTransformer, which creates an attention-based understanding of molecular graphs. The pre-training process uses 10 million unlabeled molecules for training and 1 million molecules for validation. The molecules are taken from ZINC15 \cite{sterling2015zinc} and Chembl \cite{gaulton2012chembl}. GROVER is fine-tuned on 11 benchmark datasets from MoleculeNet \cite{wu2018moleculenet} for final evaluation. \color{black} While this new architecture and self-supervised training approach offer appealing results the model uses 100M parameters, uses 250 Nvidia V100 GPUs, and takes four days for pre-training.

\paragraph{ChemRL-GEM}

Geometry Enhanced Molecular representation learning method (GEM) for Chemical Representation Learning (ChemRL) (ChemRL-GEM) draws inspiration from previous works using a graph-based approach, especially GROVER \cite{hu2019strategies, rong2020self}. ChemRL-GEM uses a large corpus of information to pre-train a model and, like GROVER, finds the ambiguity of SMILES and lack of structural information hard to build a successful model using a string-based approach \cite{fang2021chemrl}. ChemRL-GEM blames the low performance of prior graph approaches on neglecting the available molecular 3D information and improper pre-training tasks. ChemRL-GEM pre-training splits tasks into geometry-level and graph-level tasks. The geometry level tasks are again split into two types where bond length prediction, and bond angle prediction are local spatial structure predictions, and atomic distance matrices prediction is a global spatial prediction. The graph-level predictions are the Molecular ACCess System (MACCS) key prediction and the extended-connectivity fingerprint (ECFP) prediction. To build the model they designed an architecture called GeoGNN which trains on the atom-bond graph and the bond-angle graph of molecules to build a 3D structure-based understanding of the molecular graphs. ChemRL-GEM achieves SOTA performance and is an early attempt at a large 3D graph model pre-trained network. The pre-training approach uses 18 million training samples from ZINC15 and 2 million for evaluation \cite{sterling2015zinc}\color{black}. They state that pre-training a small subset of the data would take several hours using 1 Nvidia V100 GPU, and fine-tuning would require 1-2 days on the same GPU. As a rough estimate of the actual training process there was a follow-up work called LiteGEM which removed the 3D input of the model but still uses 74 million parameters and takes roughly ten days of training using 1 Nvidia V100 GPU \cite{zhang2021litegem}.

\paragraph{Galactica}

Galactica is inspired directly by previous large language models and their utilization of large datasets to pre-train models for downstream tasks \cite{devlin2018bert, radford2018improving}. Differentiating from BERT, they use a decoder-only setup from Vaswani et al. \cite{vaswani2017attention}. Unlike GROVER or ChemRL, Galactica focuses on general scientific knowledge and wishes to apply it to the entirety of the scientific domain \cite{taylor2022galactica}. The Galactica model takes several forms, but the 120 billion parameter model offers the best performance. Galactica trains over 60 million individual scientific documents and 2 million SMILES strings. Galactica is trained with samples from MoleculeNet, where the molecular properties are converted to text prompts and responses. \color{black} Galactica acknowledges using SMILES they receive reduced performance gains as their model size increases, but they state this could be overcome with more samples. Galactica offers a competitive performance to graph-based approaches while offering a simplified architecture design. Unfortunately, the model requires 120 billion parameters and trains using 128 Nvidia A100 80GB nodes. Despite the massive model size, it is not SOTA for a single SMILES metric. Galactica states they need additional samples/fine-tuning to obtain SOTA results.\color{black}

\vspace{-4mm}
\section{Methods}


\subsection{Data Pre-processing} \label{pre-processing}
\vspace{-2mm}
The available MoleculeNet benchmark \cite{wu2018moleculenet} uses SMILES for its molecular representation. After reviewing some of the molecule strings, not all are canonical. Including non-canonical SMILES is problematic as SMILES grammar is already complex; the molecules are converted to RDKit's canonical form to reduce complexity. The next issue is caused by RNNs, one of the many advantages of RNN is the allowance of variable length inputs to account for a variable length of history. This is only true theoretically; in practice, RNN memory has limits, which is the focus of many newer works \cite{hochreiter1997long}. Despite this limitation, it has been recently shown that RNNs can handle input lengths of around 45-50 before the performance begins to degrade \cite{yao2019novel, yin2017comparative}. Using this knowledge, we set a maximum SMILES length of 46 for the molecules. The limitation keeps a minor majority of the molecules while allowing us to ensure the RNN is performing well. After limiting the SMILES molecular length, the SMILES are converted to SELFIES. The intention of converting SMILES to SELFIES is to reduce the grammar complexity and simplify the learning process of the RNN. SELFIES converts each element and structural component, such as rings or branches, into their label. These labels are then encoded into a numerical value based on their dictionary index. 



\vspace{-4mm}
\subsection{RNN Implementation}
\begin{figure}[t]
\vspace{-2mm}
  \centering
  \includegraphics[width=.75\linewidth]{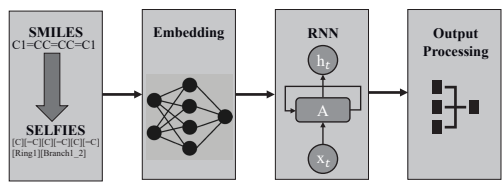}
  \caption{Overview of the RNN process. }   \label{fig:process_overview}
  \vspace{-8mm}
\end{figure}

\vspace{-2mm}
 Fig. ~\ref{fig:process_overview} offers a visualization of the methodology used to train the RNN. The molecules are first loaded in from a dataset from the MoleculeNet benchmark \cite{wu2018moleculenet} and converted to SELFIES representation using the method described in Section ~\ref{pre-processing}. The converted SELFIES are then processed through an embedding layer with a dimensional space matching the size of the label dictionary. The dictionary consists of all the unique SELFIES components within the dataset and the embedding dimension equals the dictionary size to maintain as much information as possible. The input, hidden, and output dimensions of the RNN are also equal to the size of the dictionary. Maintaining the dimensional space and not reducing it before output generation gives the RNN a chance of learning the molecular context. Fig. ~\ref{fig:RNN} and Fig. ~\ref{fig:GRU} {are visualizations of the RNN architectures used to process the SELFIES.} \color{black} RNNs historically use the Tanh activation function, but we use the LeakyReLU as it reduces saturation possibilities and typically results in higher performance \cite{maas2013rectifier, xu2015empirical}. In addition to this, we also include a dropout layer on the output of the RNN which helps prevent overfitting and reduce the error rate of RNNs \cite{gal2016theoretically}. After processing the SELFIES through the RNN, the final state should have all important prior information encoded into it. This vector then passes through an additional LeakyReLU and dropout layer before being fed to a fully connected layer. The fully connected layer reduces the vector from the dictionary-sized dimension down to the number of classes present in the molecular property. Subsequently, a soft-max operation finds the most likely class.

\vspace{-4mm}
\section{Results and Comparative Analysis}

\subsection{Results}
\begin{figure}[t]
\vspace{-2mm}
    \centering
    \includegraphics[width=\linewidth]{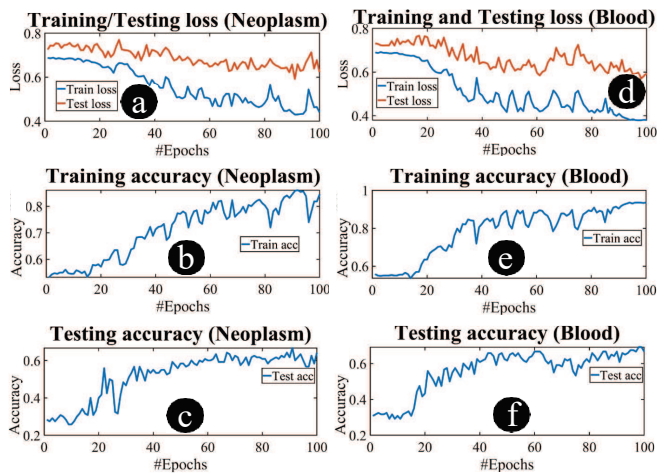}
    \caption{Results of three tasks: (a) loss curves, (b) training accuracy, (c) testing accuracy for neoplasms benign, malignant and unspecified (incl cysts and polyps) disorders, (d), (e), (f) for blood and lymphatic system disorders.} 
    \label{fig:balanced_3_plots}
    \vspace{-6mm}
\end{figure}

\vspace{-2mm}
Before training on the selected MoleculeNet datasets referenced in Section \ref{molnet_background}\color{black}, we perform an additional reduction to the dataset by setting the lower bound of 31 molecules to the SMILES string allowing for the search space to remain sufficiently complex while reducing the overall run time. The lower bound reduces the datasets before stratified splitting the data using 80\% for training and 20\% for testing \cite{thompson2012sampling}. The stratified splitting intends to maintain the known sample rate of a given side effect to model real-world testing. However, during training, we want to remove the sampling bias to ensure our model accurately learns the causes of a side effect. The minority samples within the training set are duplicated to have an even sample count between the side effect present and the side effect not present to reduce the sampling bias. After replicating training samples, the SMILES conversion to SELFIES occurs. Typical natural language processing (NLP) methods use a word, sub-word, or character tokenization to convert strings into numerical values, but we opt for a slightly different method, which we explain by referring to equation ~\ref{eq:benzene}. It shows the SELFIES representation of benzene where each molecule and structural element are between brackets. Using this representation, we decide to tokenize based on each set of brackets that exist within the SELFIES converted dataset. This results in a total of 47 unique values. After tokenizing the SELFIES, the embedding dimension, input dimension of the RNN, and the hidden dimension of the RNN are set to a size of 47 to match the dimensional space of the tokens. To give the RNN model the best opportunity to make accurate classifications, we use a single model to perform a single side effect classification prediction. For SIDER, instead of predicting all 27 potential side effect classifications, we opt to predict 20 side effect classifications due to extreme imbalances present in the side effect data. The vanilla RNN architecture results in a model with 11.5K parameters and the GRU architecture results in a model with 18.8k parameters. Both train in under 2 minutes on an Nvidia GeForce RTX 3090. To compare our performance with other works that use MoleculeNet we evaluate using {the suggested metric,} the receiver operating characteristic curve (ROC) \cite{yang2019analyzing, zhou2023uni}. 
While ROC is helpful for comparison, it is commonly misunderstood \cite{doi:10.1148/radiology.143.1.7063747, saito2015precision} so we include a small sample of 2 training/testing accuracy and loss curves in Fig. ~\ref{fig:balanced_3_plots} as a simple spot check of model performance. Examining Fig. ~\ref{fig:balanced_3_plots}, we note that training and testing loss is decreasing across all three side effect properties. There are spikes within each of the loss curves, but this is known to have occurred since the inception of RNNs \cite{elman1990finding}. The training loss for all three side effects saturate faster than the testing. There can be some gap in performance in loss based on the difficulty of new samples, but the gap here is likely accentuated as an unfortunate side effect of the minority sample duplication process. The duplicate samples within the training set help the model learn what molecular components help detect a side effect, but during training, the repeated samples become easier to predict for the model. In the case of accuracy, both training and testing show an upward trending curve where improvement starts to attenuate between the 20th-40th epoch. This attenuation roughly matches the attenuation that occurs with the loss curves. Comparing training and testing accuracy there appears to be a roughly 20+\% gap in performance at nearly every epoch, which we again attribute to the duplicate samples within the training set.

\vspace{-4mm}
\begin{equation}
[C][=C][C][=C][C][=C][Ring1][=Branch1] \label{eq:benzene}
\end{equation}
\vspace{-8mm}

\vspace{-2mm}
\subsection{Comparisons}
\vspace{-2mm}
To understand the performance of the proposed approach, we compare it across multiple datasets to two top-performing GNN models, ChemRL-GEM \cite{fang2021chemrl} and GROVER$_{large}$ \cite{rong2020self}, and a top-performing NLP model, Galactica \cite{taylor2022galactica}. In addition to the top-performing large models, we include the random forest and GCN model from MoleculeNet \cite{wu2018moleculenet}, the DMPNN model \cite{yang2019analyzing}, and the pre-trained GIN model \cite{hu2019strategies}. For each heuristic model we evaluate we train using 20 different random seeds and evaluate the model by taking the top 3 performing ROC scores per metric. We include standard deviation as a way to account for uncertainty in model performance. Overall results are shown in Table \ref{tab:OVERALL_ROC}. Beginning with the SIDER test, the results in Table ~\ref{tab:OVERALL_ROC} show our approach using the GRU achieves SOTA performance with a 17.8\% higher performance over the best model not using the proposed method (RF \cite{wu2018moleculenet}). While there are no direct statistics available for ChemRL-GEM, we use roughly 99.7\% fewer parameters than its follow-up work, LiteGEM \cite{fang2021chemrl, zhang2021litegem}. It is worth noting that applying our proposed approach to a CNN network offers a 17.3\% higher performance over RF \cite{wu2018moleculenet}. For the BBBP test, the GRU outperforms ChemRL-GEM and Galactica but performs 0.32\% worse than GROVER$_{large}$. While it may be possible that GROVER achieves better performance due to their usage of graph representation, it more likely stems from having 100M parameters, over 5,000x more parameters than our model \cite{rong2020self}. For the Clintox test, our performance was again the best of all the models. This test is one of Galactica's best performances, yet we can achieve a 17.05\% higher performance with 6Mx less parameters \cite{taylor2022galactica}. Comparing the GRU approach to the best performing approach for Clintox and BACE, GROVER$_{large}$ \cite{rong2020self}, it achieves a 3.74\% better performance for Clintox and a 5.01\% better performance for BACE.

Further comparing the models we include the pre-train data requirement, the train time of the model and the GPU requirement to achieve the listed runtime in Table \ref{tab:OVERALL_ROC}.
\color{black}

\begin{table*}[t]
  \caption{Table of ROC performance per molecular property prediction across datasets.}
  \label{tab:OVERALL_ROC}
  \centering
  \vspace{-2mm}
  \begin{adjustbox}{width= \linewidth}
        \begin{tabular}{|c|c|c|c|c|c|c|c|c|c|c|c|c|}        
            \hline
            \textbf{Model} & \textbf{RNN} & \textbf{bi-RNN} & \textbf{GRU} & \textbf{CNN} & \textbf{MLP} & ChemRL-GEM \cite{fang2021chemrl} & GROVER$_{large}$ \cite{rong2020self} & Galactica \cite{taylor2022galactica} & RF \cite{wu2018moleculenet} & GCN \cite{wu2018moleculenet} & DMPNN \cite{yang2019analyzing} & Pre-trained GIN \cite{hu2019strategies} \\
            \hline
            SIDER & \textit{.557 $\pm$ .05 } & \textit{.56 $\pm$ .071 } & \textbf{.818 $\pm$ .084} & \textbf{.814 $\pm$ .059} & \textbf{.731 $\pm$ .063} & .672 $\pm$ .004 & .658 $\pm$ .023 & .632 & .684 $\pm$ .009 & .638 $\pm$ .012 & .676 $\pm$ .014 & .627 $\pm$ .08 \\
            \hline
            BBBP & \textit{.643 $\pm$ .133} & \textit{.5 $\pm$ 0} & \textit{.937 $\pm$ .025} & \textit{.923 $\pm$ .068} & \textit{.681 $\pm$ .039} & .724 $\pm$ .004 & .940 $\pm$ .019 & .661 & .714 $\pm$ .000 & .690 $\pm$ .009 & .737 $\pm$ .001 & .687 $\pm$ .013 \\
            \hline
            Clintox & \textit{.637 $\pm$ .101 } & \textit{.6 $\pm$ .17} & \textbf{.98 $\pm$ .012} & \textbf{.976 $\pm$ .021} & \textit{.860 $\pm$ .029} & .901 $\pm$ .013 & .944 $\pm$ .021 & .826 & .713 $\pm$ .056 & .807 $\pm$ .047 & .864 $\pm$ .017 & .726 $\pm$ .015 \\
            \hline
            BACE & \textit{.623 $\pm$ .015 } & \textit{.733 $\pm$ .196} & \textbf{.94 $\pm$ .006} & \textbf{.923 $\pm$ .006} & \textbf{.943 $\pm$ .017} & .856 $\pm$ .011 & .894 $\pm$ .028 & .617 & .867 $\pm$ .008 & .783 $\pm$ .014 & .852 $\pm$ .053 & .845 $\pm$ .007 \\
            \hline
            HIV & \textit{.583 $\pm$ .032} & \textit{.61 $\pm$ .095} & \textbf{.653 $\pm$ .031} & \textbf{.713 $\pm$ .081} & \textbf{.683 $\pm$ .041} & - & - & .632 & - & .763 $\pm$ .016 & .776 $\pm$ .007 & .799 $\pm$ .007 \\
            \hline
            MUV & \textit{.621 $\pm$ .089} & \textit{.627 $\pm$ .107} & \textit{.923 $\pm$ .034} & \textit{.931 $\pm$ .037} & \textit{.867 $\pm$ .058} & - & - & - & - & .046 $\pm$ .031 & .041 $\pm$ .007 & .813 $\pm$ .021 \\
            \hline
            Pre-train data & - & - & - & - & - & 18M Mols & 10M Mols & 60M Docs & - & - & - & 2.4M Mols \\
            \hline
            Rep/Est Train Time & $\sim$ 2m & $\sim$ 2m & $\sim$ 2m & $\sim$ 2m & $\sim$ 2m & $\sim$ 10D & 4D & 30D & - & - & - & - \\
            \hline
            GPU Req. & RTX3090 & RTX3090 & RTX3090 & RTX3090 & RTX3090 & $\sim$ 1 V100 & 250 V100 & 512 A100 & - & - & - & - \\
            \hline
            
        \end{tabular}
        \end{adjustbox}
  \vspace{-6mm}
\end{table*}

\vspace{-4mm}
\section{Discussion \& Conclusion}
\vspace{-2mm}
While large data models may offer coverage of larger molecular lengths, we have shown that small models (specifically GRUs) are still viable candidates for molecular property prediction. Smaller models are cheaper, more practical, and more accessible solutions as they don't require multiple GPUs and several days of training prior to obtaining a result. There are limitations with RNN based models, but when these limitations are carefully considered and more descriptive languages, such as SELFIES \cite{krenn2020self}, are used RNNs offer SOTA/near SOTA results.

\vspace{-4mm}
\subsection{Clinical Insights}
\vspace{-2mm}
Property prediction models allow chemists to perform molecular evaluations prior to physical experimentation. Effective property evaluation can prevent months of wet lab research being spent on molecules that will not be feasible. Reducing failures realized during synthesis has the potential to greatly reduce the drug to market run time, enabling clinical researchers to rapidly treat patients for their medical conditions.
\color{black}

\vspace{-4mm}
\subsection{Constraints}
\vspace{-2mm}
Despite the RNN's learning capabilities and ability to process variable length input with no additional parameters required, using such an architecture does have drawbacks. RNN models can scale when dealing with large datasets via batching or even model parallelization, but they do not scale well when considering larger input sequences. Theoretically, RNN models can process large sequences of information with no problem, but in practice, RNNs can suffer from vanishing or exploding gradients causing them to "forget" important information. Even if we could implement the perfect memory model, the RNN still suffers from long run times where each addition to the sequence increases the run time due to the sequential nature of recurrence. One possible method to mitigate the long run time would be chunking, where the sequences are partitioned into smaller processable pieces. Unfortunately, this is unreliable, as sometimes vital state information may be separated from chunks causing inaccurate results.
\color{black}

\vspace{-4mm}
\subsection{Ethical Statement}
\vspace{-2mm}
While machine learning models can help identify potential molecular properties, they are not without flaws. Even if machine learning models can accurately identify all molecular properties of the datasets, they are trained with are fully dependent on previous human discoveries. The datasets are subject to flawed understandings of chemistry and even political choices. For example, the NIH only classifies drugs as toxic to the liver after successfully ruling out other potential causes \footnote{https://www.ncbi.nlm.nih.gov/books/NBK548049/}. Therefore, machine learning models should only be used for preliminary evaluation of molecules and not as the only form of molecular evaluation.
\color{black}


\textbf{Acknowledgments:}
The work is supported in parts by NSF (CNS-1722557, CNS-2129675, CCF-2210963, CCF-1718474, OIA-2040667, DGE-1723687, DGE-1821766 and DGE-2113839) and gifts from Intel.


\vspace{-4mm}
\bibliographystyle{IEEEtran}

\bibliography{test}

\begin{thebibliography}{10}
\providecommand{\url}[1]{#1}
\csname url@samestyle\endcsname
\providecommand{\newblock}{\relax}
\providecommand{\bibinfo}[2]{#2}
\providecommand{\BIBentrySTDinterwordspacing}{\spaceskip=0pt\relax}
\providecommand{\BIBentryALTinterwordstretchfactor}{4}
\providecommand{\BIBentryALTinterwordspacing}{\spaceskip=\fontdimen2\font plus
\BIBentryALTinterwordstretchfactor\fontdimen3\font minus \fontdimen4\font\relax}
\providecommand{\BIBforeignlanguage}[2]{{%
\expandafter\ifx\csname l@#1\endcsname\relax
\typeout{** WARNING: IEEEtran.bst: No hyphenation pattern has been}%
\typeout{** loaded for the language `#1'. Using the pattern for}%
\typeout{** the default language instead.}%
\else
\language=\csname l@#1\endcsname
\fi
#2}}
\providecommand{\BIBdecl}{\relax}
\BIBdecl

\bibitem{yang2019analyzing}
K.~Yang \emph{et~al.}, ``Analyzing learned molecular representations for property prediction,'' \emph{Journal of chemical information and modeling}, vol.~59, no.~8, pp. 3370--3388, 2019.

\bibitem{wieder2020compact}
O.~Wieder \emph{et~al.}, ``A compact review of molecular property prediction with graph neural networks,'' \emph{Drug Discovery Today: Technologies}, vol.~37, pp. 1--12, 2020.

\bibitem{lengauer2004novel}
T.~Lengauer \emph{et~al.}, ``Novel technologies for virtual screening,'' \emph{Drug discovery today}, vol.~9, no.~1, pp. 27--34, 2004.

\bibitem{merkwirth2005automatic}
C.~Merkwirth \emph{et~al.}, ``Automatic generation of complementary descriptors with molecular graph networks,'' \emph{Journal of chemical information and modeling}, vol.~45, no.~5, pp. 1159--1168, 2005.

\bibitem{hu2019strategies}
W.~Hu \emph{et~al.}, ``Strategies for pre-training graph neural networks,'' \emph{arXiv preprint arXiv:1905.12265}, 2019.

\bibitem{wu2020comprehensive}
Z.~Wu \emph{et~al.}, ``A comprehensive survey on graph neural networks,'' \emph{IEEE transactions on neural networks and learning systems}, vol.~32, no.~1, pp. 4--24, 2020.

\bibitem{mayr2018large}
Mayr \emph{et~al.}, ``Large-scale comparison of machine learning methods for drug target prediction on chembl,'' \emph{Chemical science}, vol.~9, no.~24, pp. 5441--5451, 2018.

\bibitem{taylor2022galactica}
R.~Taylor \emph{et~al.}, ``Galactica: A large language model for science,'' \emph{arXiv preprint arXiv:2211.09085}, 2022.

\bibitem{wu2018moleculenet}
Wu \emph{et~al.}, ``Moleculenet: a benchmark for molecular machine learning,'' \emph{Chemical science}, vol.~9, no.~2, pp. 513--530, 2018.

\bibitem{kuhn2016sider}
M.~Kuhn \emph{et~al.}, ``The sider database of drugs and side effects,'' \emph{Nucleic acids research}, vol.~44, no.~D1, pp. D1075--D1079, 2016.

\bibitem{subramanian2016computational}
Subramanian \emph{et~al.}, ``Computational modeling of $\beta$-secretase 1 (bace-1) inhibitors using ligand based approaches,'' \emph{Journal of chemical information and modeling}, vol.~56, no.~10, pp. 1936--1949, 2016.

\bibitem{martins2012bayesian}
Martins \emph{et~al.}, ``A bayesian approach to in silico blood-brain barrier penetration modeling,'' \emph{Journal of chemical information and modeling}, vol.~52, no.~6, pp. 1686--1697, 2012.

\bibitem{rohrer2009maximum}
Rohrer \emph{et~al.}, ``Maximum unbiased validation (muv) data sets for virtual screening based on pubchem bioactivity data,'' \emph{Journal of chemical information and modeling}, vol.~49, no.~2, pp. 169--184, 2009.

\bibitem{weininger1988smiles}
D.~Weininger, ``Smiles, a chemical language and information system. 1. introduction to methodology and encoding rules,'' \emph{Journal of chemical information and computer sciences}, vol.~28, no.~1, pp. 31--36, 1988.

\bibitem{krenn2020self}
M.~Krenn \emph{et~al.}, ``Self-referencing embedded strings (selfies): A 100\% robust molecular string representation,'' \emph{Machine Learning: Science and Technology}, vol.~1, no.~4, p. 045024, 2020.

\bibitem{elman1990finding}
J.~L. Elman, ``Finding structure in time,'' \emph{Cognitive science}, vol.~14, no.~2, pp. 179--211, 1990.

\bibitem{cho2014learning}
Cho \emph{et~al.}, ``Learning phrase representations using rnn encoder-decoder for statistical machine translation,'' \emph{arXiv preprint arXiv:1406.1078}, 2014.

\bibitem{rong2020self}
Rong \emph{et~al.}, ``Self-supervised graph transformer on large-scale molecular data,'' \emph{Advances in Neural Information Processing Systems}, vol.~33, pp. 12\,559--12\,571, 2020.

\bibitem{devlin2018bert}
Devlin \emph{et~al.}, ``Bert: Pre-training of deep bidirectional transformers for language understanding,'' \emph{arXiv preprint arXiv:1810.04805}, 2018.

\bibitem{radford2018improving}
Radford \emph{et~al.}, ``Improving language understanding by generative pre-training,'' 2018.

\bibitem{wang2019smiles}
Wang \emph{et~al.}, ``Smiles-bert: large scale unsupervised pre-training for molecular property prediction,'' in \emph{Proceedings of the 10th ACM international conference on bioinformatics, computational biology and health informatics}, 2019, pp. 429--436.

\bibitem{sterling2015zinc}
Sterling \emph{et~al.}, ``Zinc 15--ligand discovery for everyone,'' \emph{Journal of chemical information and modeling}, vol.~55, no.~11, pp. 2324--2337, 2015.

\bibitem{gaulton2012chembl}
Gaulton \emph{et~al.}, ``Chembl: a large-scale bioactivity database for drug discovery,'' \emph{Nucleic acids research}, vol.~40, no.~D1, pp. D1100--D1107, 2012.

\bibitem{fang2021chemrl}
Fang \emph{et~al.}, ``Chemrl-gem: Geometry enhanced molecular representation learning for property prediction,'' \emph{arXiv preprint arXiv:2106.06130}, 2021.

\bibitem{zhang2021litegem}
Zhang \emph{et~al.}, ``Litegem: Lite geometry enhanced molecular representation learning for quantum property prediction,'' \emph{arXiv preprint arXiv:2106.14494}, 2021.

\bibitem{vaswani2017attention}
Vaswani \emph{et~al.}, ``Attention is all you need,'' \emph{Advances in neural information processing systems}, vol.~30, 2017.

\bibitem{hochreiter1997long}
Hochreiter \emph{et~al.}, ``Long short-term memory,'' \emph{Neural computation}, vol.~9, no.~8, pp. 1735--1780, 1997.

\bibitem{yao2019novel}
X.~Yao \emph{et~al.}, ``A novel independent rnn approach to classification of seizures against non-seizures,'' \emph{arXiv preprint arXiv:1903.09326}, 2019.

\bibitem{yin2017comparative}
W.~Yin \emph{et~al.}, ``Comparative study of cnn and rnn for natural language processing,'' \emph{arXiv preprint arXiv:1702.01923}, 2017.

\bibitem{maas2013rectifier}
Maas \emph{et~al.}, ``Rectifier nonlinearities improve neural network acoustic models,'' in \emph{Proc. icml}, vol.~30, no.~1.\hskip 1em plus 0.5em minus 0.4em\relax Atlanta, Georgia, USA, 2013, p.~3.

\bibitem{xu2015empirical}
Xu \emph{et~al.}, ``Empirical evaluation of rectified activations in convolutional network,'' \emph{arXiv preprint arXiv:1505.00853}, 2015.

\bibitem{gal2016theoretically}
Gal \emph{et~al.}, ``A theoretically grounded application of dropout in recurrent neural networks,'' \emph{Advances in neural information processing systems}, vol.~29, 2016.

\bibitem{thompson2012sampling}
\BIBentryALTinterwordspacing
S.~Thompson, \emph{Sampling}, ser. CourseSmart.\hskip 1em plus 0.5em minus 0.4em\relax Wiley, 2012. [Online]. Available: \url{https://books.google.com/books?id=-sFtXLIdDiIC}
\BIBentrySTDinterwordspacing

\bibitem{zhou2023uni}
Zhou \emph{et~al.}, ``Uni-mol: A universal 3d molecular representation learning framework,'' 2023.

\bibitem{doi:10.1148/radiology.143.1.7063747}
\BIBentryALTinterwordspacing
Hanley \emph{et~al.}, ``The meaning and use of the area under a receiver operating characteristic (roc) curve.'' \emph{Radiology}, vol. 143, no.~1, pp. 29--36, 1982, pMID: 7063747. [Online]. Available: \url{https://doi.org/10.1148/radiology.143.1.7063747}
\BIBentrySTDinterwordspacing

\bibitem{saito2015precision}
Saito \emph{et~al.}, ``The precision-recall plot is more informative than the roc plot when evaluating binary classifiers on imbalanced datasets,'' \emph{PloS one}, vol.~10, no.~3, p. e0118432, 2015.

\end{thebibliography}


\vspace{-4mm}
\appendix
\vspace{-2mm}
\subsection{Accuracy, Precision and Recall Statisitics}
\vspace{-2mm}

\begin{table}[htp]
    \centering
    \begin{adjustbox}{width= \linewidth}
        \begin{tabular}{|c|c|c|c|c|c|}
             \hline
             \textbf{Data} & MLP & CNN & GRU & RNN & BiRNN \\
             \hline
             SIDER & .453 $\pm$ .052 & .555 $\pm$ .039 & .598 $\pm$ .049 & .314 $\pm$ .126 & .353 $\pm$ .156 \\
             \hline
             BBBP & .346 $\pm$ .019 & .587 $\pm$ .026 & .641 $\pm$ .008 & .129 $\pm$ .004 & .331 $\pm$ .274 \\
             \hline
             ClinTox & .546 $\pm$ .031 & .797 $\pm$ .022 & .824 $\pm$ .008 & .102 $\pm$ .033 & .241 $\pm$ .250 \\
             \hline
             BACE & .695 $\pm$ .033 & .657 $\pm$ .057 & .808 $\pm$ .029 & .404 $\pm$ .174 & .174 $\pm$ .024 \\
             \hline
             HIV & .525 $\pm$ .011 & .529 $\pm$ .013 & .513 $\pm$ .022 & .476 $\pm$ .036 & .486 $\pm$ .014 \\
             \hline
             MUV & .581 $\pm$ .110 & .683 $\pm$ .083 & .478 $\pm$ .067 & .304 $\pm$ .191 & .311 $\pm$ .21 \\
             \hline
        \end{tabular}
    \end{adjustbox}    
    \caption{Model accuracy results using proposed method}
    \label{tab:my_label}
\end{table}

\vspace{-4mm}
\begin{table}[htp]
    \centering
    \begin{adjustbox}{width= \linewidth}
        \begin{tabular}{|c|c|c|c|c|c|}
            \hline
             \textbf{Data} & MLP & CNN & GRU & RNN & BiRNN \\
             \hline
             SIDER & .734 $\pm$ .317 & .773 $\pm$ .272 & .799 $\pm$ .214 & .526 $\pm$ .399 & .488 $\pm$ .413 \\
             \hline
             BBBP & .845 $\pm$ .007 & 1 $\pm$ 0 & 1 $\pm$ 0 & .0 $\pm$ 0 & .6 $\pm$ .432 \\
             \hline
             ClinTox & .295 $\pm$ .082 & .639 $\pm$ .104 & .667 $\pm$ .236 & .174 $\pm$ .029 & .083 $\pm$ .029 \\
             \hline
             BACE & .444 $\pm$ .079 & .528 $\pm$ .171 & .704 $\pm$ .127 & .248 $\pm$ .12 & .207 $\pm$ .139 \\
             \hline
             HIV & .536 $\pm$ .081 & .677 $\pm$ .087 & .806 $\pm$ .056 & .739 $\pm$ .055 & .604 $\pm$ .152 \\
             \hline
             MUV & 1 $\pm$ 0 & .495 $\pm$ .209 & .391 $\pm$ .093 & .203 $\pm$ .140 & .229 $\pm$ .186 \\
             \hline
        \end{tabular}
    \end{adjustbox}    
    \caption{Model precision results using proposed method}
    \label{tab:my_label}
\end{table}

\vspace{-4mm}
\begin{table}[htp]
    \centering
    \begin{adjustbox}{width= \linewidth}
        \begin{tabular}{|c|c|c|c|c|c|}
            \hline
             \textbf{Data} & MLP & CNN & GRU & RNN & BiRNN \\
             \hline
             SIDER & .676 $\pm$ .240 & .757 $\pm$ .189 & .739 $\pm$ .173 & .438 $\pm$ .404 & .478 $\pm$ .444 \\
             \hline
             BBBP & .845 $\pm$ .007 & .75 $\pm$ .029 & .724 $\pm$ .007 & 0 $\pm$ 0 & .364 $\pm$ .452 \\
             \hline
             ClinTox & 1 $\pm$ 0 & 1 $\pm$ 0 & 1 $\pm$ 0 & 1 $\pm$ 0 & 1 $\pm$ 0 \\
             \hline
             BACE & 1 $\pm$ 0 & 1 $\pm$ 0 & 1 $\pm$ 0 & .833 $\pm$ .236 & 1 $\pm$ 0 \\
             \hline
             HIV & .736 $\pm$ .086 & .852 $\pm$ .139 & .606 $\pm$ .048 & .516 $\pm$ .152 & .616 $\pm$ .118 \\
             \hline
             MUV & .362 $\pm$ .079 & 1 $\pm$ 0 & 1 $\pm$ 0 & .899 $\pm$ .199 & .892 $\pm$ .215 \\
             \hline
        \end{tabular}
    \end{adjustbox}    
    \caption{Model recall results using proposed method}
    \label{tab:my_label}
\end{table}

\vspace{-6mm}
\subsection{Wilcoxon Statisitics}
To understand the statistical difference between each of the models running with the proposed approach a Wilcoxon Signed-Rank test was performed for each model against the absolute best performance per dataset between the 3 SOTA models (Galactica \cite{taylor2022galactica}, ChemRL-GEM \cite{fang2021chemrl},and GROVER$_{large}$ \cite{rong2020self}).

For the GRU the Wilcoxon Signed-Rank test indicated that there is a non-significant large difference between GRU (Mdn = 0.9 ,n = 5) and Top SOTA Results (Mdn = 0.9 ,n = 5), W+ = 1, p = .125, r = -0.7. 

For the CNN the Wilcoxon Signed-Rank test indicated that there is a non-significant large difference between CNN (Mdn = 0.9 ,n = 5) and Top SOTA Results (Mdn = 0.9 ,n = 5), W+ = 1, p = .125, r = -0.7. 

For the MLP the Wilcoxon Signed-Rank test indicated that there is a non-significant small difference between MLP (Mdn = 0.7 ,n = 5) and Top SOTA Results (Mdn = 0.9 ,n = 5), W+ = 9, p = .813, r = 0.1 

For the RNN the Wilcoxon Signed-Rank test indicated that there is a non-significant large difference between RNN (Mdn = 0.6 ,n = 5) and Top SOTA Results (Mdn = 0.9 ,n = 5), W+ = 15, p = .063, r = 0.8. 

For the bidirectional RNN the Wilcoxon Signed-Rank test indicated that there is a non-significant large difference between Bidirectional RNN (Mdn = 0.6 ,n = 5) and Top SOTA Results (Mdn = 0.9 ,n = 5), W+ = 15, p = .063, r = 0.8. 

In addition to the 3 SOTA models we also performed a Wilcoxon Signed-Rank test comparing the GRU vs MLP and the GRU vs CNN.

Comparing the GRU vs MLP the Wilcoxon Signed-Rank test indicated that there is a non-significant large difference between GRU (Mdn = 0.9 ,n = 6) and MLP (Mdn = 0.8 ,n = 6), W+ = 3, p = .156, r = -0.6. 

Comparing the GRU vs CNN the Wilcoxon Signed-Rank test indicated that there is a non-significant very small difference between GRU (Mdn = 0.9 ,n = 6) and CNN (Mdn = 0.9 ,n = 6), Z = -0.2, p = .833, r = -0.09.

\newpage

\vfill

\end{document}